\documentstyle[aps,twocolumn]{revtex}
\input{epsf}

\title{Quantum information and precision measurement\thanks{CALT-68-2217}}

\author{Andrew M. Childs\thanks{\tt amchilds@caltech.edu}, John
Preskill,\thanks{\tt preskill@theory.caltech.edu}and Joseph Renes\thanks{\tt
renes@its.caltech.edu}\\
{\sl Lauritsen Laboratory of High Energy Physics\\
California Institute of Technology\\
Pasadena, CA 91125, USA}}
\date{April, 1999}

\begin{document}

\maketitle

\begin{abstract}
We describe some applications of quantum information theory to the analysis of
quantum limits on measurement sensitivity. A measurement of a weak force acting
on a quantum system is a determination of a classical parameter appearing in
the master equation that governs the evolution of the system; limitations on
measurement accuracy arise because it is not possible to distinguish perfectly
among the different possible values of this parameter.
Tools developed in the study of quantum information and computation can be
exploited to improve the precision of physics experiments; examples include
superdense coding, fast database search, and the quantum Fourier transform.

\end{abstract}


\section{Introduction: Distinguishability of superoperators}
The exciting recent developments in the theory of quantum information and
computation have already established an enduring legacy.  The two most
far-reaching results --- that a quantum computer (apparently) can solve
problems that will forever be beyond the reach of classical computers
\cite{qc}, and that quantum information can be protected from errors if
properly encoded \cite{qec} --- have surely earned a prominent place at the
foundations of computer science.

The implications of these ideas for the future of physics are less clear, but
we expect them to be profound. In particular, we anticipate that our deepening
understanding of quantum information will lead to new strategies for pushing
back the boundaries of quantum-limited measurements.  Quantum entanglement,
quantum error correction, and quantum information processing can all be
exploited to improve the information-gathering capability of physics
experiments.

In a typical quantum-limited measurement, a classical signal is conveyed over a
quantum channel \cite{mabuchi}. Nature sends us a message, such as the value of
a weak force, that can be regarded as a classical parameter appearing in the
Hamiltonian of the apparatus (or more properly, if there is noise, its master
equation).  The apparatus undergoes a quantum operation $\$(a)$, and we are to
extract as much information as we can about the parameter(s) $a$ by choosing an
initial preparation of the apparatus, and a positive-operator-valued measure
(POVM) to read it out.  Quantum information theory should be able to provide a
theory of the {\sl distinguishability of superoperators}, a measure of how much
information we can extract that distinguishes one superoperator from another,
given some specified resources that are available for the purpose.  This
distinguishability measure would characterize the inviolable limits on
measurement precision that can be achieved with fixed resources.

Many applications of quantum information theory involve the problem of
distinguishing nonorthogonal quantum {\sl states}.  For example, a density
operator $\rho_a $ is chosen at random from an ensemble ${\cal
E}=\{\rho_a,p_a\}$ (where $p_a$ is an {\it a priori} probability), and a
measurement is performed to extract information about which $\rho_a$ was
chosen.  The problem of distinguishing {\sl superoperators} is rather
different, but the two problems are related.  For example, let us at first
ignore noise, and also suppose that the classical force we are trying to detect
is static.  Then we are trying to identify a particular time-independent
Hamiltonian $H_a$ that has been drawn from an ensemble $\{H_a,p_a\}$. We may
choose a particular initial pure state $|\psi_0\rangle$, and then allow the
state to evolve, as governed by the unknown Hamiltonian, for a time $t$; our
ensemble of possible Hamiltonians generates an ensemble of pure states
\begin{equation}
\{|\psi_a(t)\rangle = e^{-itH_a}|\psi_0\rangle, p_a\}~.
\end{equation}
Since our goal is to gain as much information as possible about the applied
Hamiltonian, we should choose the initial state $|\psi_0\rangle$ so that the
resulting final states are maximally distinguishable.

There are many variations on the problem, distinguished in part by the
resources we regard as most valuable.  We might have the freedom to chose the
elapsed time as we please, or we might impose constraints on $t$.  We might
have the freedom to modify the Hamiltonian by adding an additional ``driving''
term that is under our control.  We might use an {\sl adaptive} strategy, where
we make repeated (possibly weak) measurements, and our choice of initial state
or driving term in later measurements takes into account the information
already collected in earlier measurements \cite{wiseman_adaptive}.

Imposing an appropriate cost function on resources is an important aspect of
the formulation of the problem, particularly in the case of the detection of a
static (DC) signal.  For example, we could in principle repeat the measurement
procedure many times to continually improve the accuracy of our estimate.  In
this respect, the problem of distinguishing superoperators does not have quite
so fundamental a character as the problem of distinguishing states, as in the
latter case the no-cloning principle \cite{no_clone} prevents us from making
repeated measurements on multiple copies of the unknown state.  But for a
time-dependent signal that stays ``on'' for a finite duration, there will be a
well-defined notion of the optimal strategy for distinguishing one possible
signal from another, once our apparatus and its coupling to the classical
signal have been specified.  Still, for the sake of simplicity, we will mostly
confine our attention here to the case of DC signals.

We don't know exactly what shape this nascent theory of the distinguishability
of  superoperators should take, but we hope that further research can promote
the development of new strategies for performing high-precision measurements.
On the one hand we envision a program of research that will be relevant to real
laboratory situations.  On the other hand, we seek results that are to some
degree robust and general (not tied to some particular model of decoherence, or
to a particular type of coupling between quantum probe and classical signal).
Naturally, there is some tension between these two central desiderata;  rather
than focus on a specific experimental context, we lean here toward more
abstract formulations of the problem.

Our discussion is far from definitive; its goal is to invite a broader
community to consider these issues.  We will mostly be content to observe that
some familiar concepts from the theory of quantum information and computation
can be translated into tools for the measurement of classical forces. Some
examples include superdense coding, fast database search, and the quantum
Fourier transform.

Naturally, the connections between quantum information theory and precision
measurement have been recognized previously by many authors. Especially
relevant is the work by Wootters \cite{wootters}, by Braunstein\cite{braun},
and by Braunstein and Caves \cite{braunstein} on state distinguishability and
parameter estimation, and by Braginsky and others \cite{braginsky} on quantum
nondemolition measurement. Though what we have to add may be relatively modest,
we hope that it may lead to further progress.

\section{Superdense coding: improved distinguishability through entanglement}
\label{sec:superdense}

Recurring themes of quantum information theory are that entanglement can be a
valuable resource, and that entangled measurements sometimes can collect more
information than unentangled measurements.  It should not be surprising, then,
if the experimental physicist finds that the best strategies for detecting a
weak classical signal involve the preparation of entangled states and the
measurement of entangled observables.

Suppose, for example, that our apparatus is a single-qubit, whose
time-independent Hamiltonian (aside from an irrelevant additive constant), can
be expressed as
\begin{equation}
H_{\vec a}=\vec a\cdot \vec \sigma~;
\end{equation}
here $\vec a=(a_1,a_2,a_3)$ is an unknown three-vector, and $\sigma_{1,2,3}$
are the Pauli matrices.  (We may imagine that a spin-${1\over 2}$ particle with
a magnetic moment is employed to measure a static magnetic field.) By preparing
an initial state of the qubit, allowing the qubit to evolve, and then
performing a single measurement, we can extract at best one bit of information
about the magnetic field (as Holevo's theorem \cite{holevo} ensures that the
optimal POVM in a two-dimensional Hilbert space can acquire at most one bit of
information about a quantum state).

If we have two qubits, and measure them one at a time, we can collect at best
two bits of information about the magnetic field.  In principle, this could be
enough to distinguish perfectly among four possible values of the field.  In
practice, for a generic choice of four Hamiltonians labeled by vectors $\vec
a^{(1,2,3,4)}$, the optimal information gain cannot be achieved by measuring
the qubits one at a time. Rather a better strategy exploits quantum
entanglement.

An improved strategy can be formulated by following the paradigm of superdense
coding \cite{wiesner}, whereby shared entanglement is exploited to enhance
classical communication between two parties.  To implement superdense coding,
the sender (Alice) and the receiver (Bob) use a shared Bell state
\begin{equation}
|\phi^+\rangle= {1\over\sqrt{2}}\left(|00\rangle + |11\rangle\right)
\end{equation}
that they have prepared previously.  Alice applies one of the four unitary
operators $\{I,\sigma_1,\sigma_2,\sigma_3\}$ to her member of the entangled
pair, and then sends it to Bob.  Upon receipt, Bob possesses one of the four
mutually orthogonal Bell states\begin{eqnarray}
\label{bellbasis}
|\phi^+\rangle & = & {1\over\sqrt{2}}\left(|00\rangle+|11\rangle\right) = I
\otimes I |\phi^+\rangle ~,\cr
|\psi^+\rangle & = & {1\over\sqrt{2}}\left(|01\rangle+|10\rangle\right) =
\sigma_1  \otimes I |\phi^+\rangle ~,\cr
-i|\psi^-\rangle & = & {-i\over\sqrt{2}}\left(|01\rangle-|10\rangle\right) =
\sigma_2  \otimes I |\phi^+\rangle ~,\cr
|\phi^-\rangle & = & {1\over\sqrt{2}}\left(|00\rangle-|11\rangle\right) =
\sigma_3  \otimes I |\phi^+\rangle~;\end{eqnarray}
by performing an entangled Bell measurement (simultaneous measurements of the
commuting collective observables $\sigma_1\otimes\sigma_1$ and
$\sigma_3\otimes\sigma_3$), Bob can perfectly distinguish the states. Although
only one qubit passes from Alice to Bob, two classical bits of information are
transmitted and successfully decoded.  In fact, this enhancement of the
transmission rate is optimal -- with shared entanglement, no more than two
classical bits can be carried by each transmitted qubit \cite{hausladen}.

The lesson of superdense coding is that entanglement can allow us to better
distinguish
operations on quantum states, and we may apply this method to the problem of
distinguishing Hamiltonians.\footnote{This idea was suggested to us by Chris
Fuchs \cite{fuchs_private}.}  Let us imagine that the magnitude of the magnetic
field is known, but not its direction -- then we can choose our unit of time so
that $|\vec a|=1$. We may prepare a pair of qubits in the entangled state
$|\phi^+\rangle$, and expose only one member of the pair to the magnetic field
while the other remains well shielded.  In time $t$, the state evolves to
\begin{eqnarray}
|\psi_{\hat a}(t)\rangle & \equiv &\exp\left(-itH_{\hat a}\otimes
I\right)|\phi^+\rangle \cr
& = & \left[\cos t(I\otimes I) -i \sin t(\hat a\cdot\vec\sigma\otimes I)
\right]|\phi^+\rangle\cr
& = & \cos t |\phi^+\rangle \cr
&&-i\sin t
\left[a_1|\psi^+\rangle -ia_2|\psi^-\rangle + a_3|\phi^+\rangle\right]~;
\end{eqnarray}
the inner product between the states arising from Hamiltonians $H_{\hat a}$ and
$H_{\hat b}$ becomes
\begin{equation}
\label{superip}
\langle\psi_{\hat a}(t)|\psi_{\hat b}(t)\rangle = \cos^2 t + (\hat a \cdot \hat
b) \sin^2 t~.
\end{equation}
For these states to be orthogonal, we require
\begin{equation}
\hat a \cdot \hat b = -\cot^2 t~.
\end{equation}
Since $\cot^2 t \ge 0$, the states are not orthogonal for any value of $t$
unless the two magnetic field directions $\hat a$ and $\hat b$ are separated by
at least $90^\circ$.

Now suppose that the magnetic field (of known magnitude) points in one of three
directions that are related by three-fold rotational symmetry.  These
directions could form a planar trine with
$\hat a\cdot \hat b=\hat a\cdot \hat c= \hat b\cdot \hat c = -1/2$, or a
``lifted trine'' with angle $\theta$ between each pair of directions, where
$-1/2\le \cos\theta \le 0$.  For any such trine of field directions, we may
evolve for a time $t$ such that
\begin{equation}
\cot^2 t = -\cos\theta ~,
\end{equation}
and perform an (entangled) orthogonal measurement to determine the field. At
the point of tetrahedral symmetry, $\cos\theta=-1/3$, we may add a fourth field
direction such that
the inner product for each pair of field directions is $-1/3$; then all four
directions can be perfectly distinguished by Bell measurement.

In this case of four field directions with tetrahedral symmetry, the two-bit
measurement outcome achieves a two-bit information gain, if the four directions
were equally likely {\it a priori}.  In contrast, no adaptive strategy in which
single qubits are measured one at a time can attain a two-bit information gain.
 This separation between the information gain attainable through entangled
measurement and that attainable through adaptive nonentangled measurement, for
the problem of distinguishing Hamiltonians, recalls the analogous separation
noted by Peres and Wootters \cite{peres} for the problem of distinguishing
nonorthogonal states.

\section{Grover's database search: improved distinguishability through driving}

Another instructive example is Grover's method \cite{grover} for searching an
unsorted database, which (as formulated by Farhi and Gutmann \cite{farhi}) we
may interpret as a method for improving the distinguishability of a set of
Hamiltonians by adding a controlled driving term.

Consider an $N$-dimensional Hilbert space with orthonormal basis
$\{|x\rangle\}, ~x=0,1,2,\dots,N-1$, and suppose that the Hamiltonian  for this
system is known to be one of the $N$ operators
\begin{equation}
H_x=E|x\rangle\langle x| ~.
\end{equation}
We are to perform an experiment that will allow us to estimate the value of
$x$.

We could, for example, prepare the initial state ${1\over \sqrt{2}}(|y\rangle +
|y'\rangle)$, allow the system to evolve for a time $T=\pi/E$, and then perform
an orthogonal measurement in the basis $|\pm\rangle={1\over \sqrt{2}}(|y\rangle
\pm |y'\rangle)$. Then we will obtain the outcome $|-\rangle$ if and only if
one of  $y,y'$ is $x$.  Searching for $x$ by this method, we would have to
repeat the experiment for  O($N$) distinct initial states to have any
reasonable chance of successfully inferring the value of $x$.

Our task becomes easier if we are able to modify the Hamiltonian by adding a
term that we control to drive the system. We choose the driving term to be
\begin{equation}
\label{grover_drive}
H_D=E|s\rangle\langle s|~,
\end{equation}
where $|s\rangle$ denotes the state
\begin{equation}
|s\rangle={1\over \sqrt{N}}\sum_{y=0}^{N-1}|y\rangle~.
\end{equation}
Then the full Hamiltonian is
\begin{equation}
H'_x=H_x+H_D=E(|x\rangle\langle x| + |s\rangle\langle s|)~,
\end{equation}
and we can readily verify that the vectors
\begin{equation}
|E_{\pm}\rangle\equiv |s\rangle \pm |x\rangle
\end{equation}
are (unconventionally normalized!) eigenstates of $H$ with the eigenvalues
\begin{equation}
E_{\pm}=E\left(1\pm {1\over\sqrt{N}}\right)~.
\end{equation}
We may prepare the initial state
\begin{equation}
|s\rangle= {1\over 2}(|E_+\rangle + |E_-\rangle)~;
\end{equation}
since the energy splitting is $\Delta E=2E/\sqrt{N}$, after a time
\begin{equation}
T=\pi/\Delta E= \pi\sqrt{N}/2E~,
\end{equation}
this state flops to the state
\begin{equation}
 {1\over 2}(|E+\rangle - |E-\rangle)=|x\rangle~.
\end{equation}
Thus, by performing an orthogonal measurement, we can learn the value of $x$
with certainty \cite{farhi}.

The driving term we have chosen is the continuous time analog of the iteration
employed by Grover \cite{grover} for rapid searching.  And as the Grover search
algorithm can be seen to be optimal, in the sense that a marked state can be
identified with high probability with the minimal number of oracle calls
\cite{bbbv}, so the driving term we have chosen is optimal in the sense that it
enables us to identify the value of the classical parameter labeling the
Hamiltonian in the minimal time, at least asymptotically for $N$ large. (In a
physics experiment, the ``oracle'' is Nature, whose secrets we are eager to
expose.) For this Grover-Farhi-Gutmann problem, we can make a definite
statement about how to optimize expenditure of a valuable resource (time) in
the identification of a system Hamiltonian.

We also note that adding a driving term can sometimes improve the efficacy of
the superdense coding method described in \S\ref{sec:superdense}. For example,
in the case of three magnetic fields of equal magnitude with threefold
symmetry, but with an
angle between fields of less than $90^\circ$, applying a driving field along
the line of
symmetry can make the resultant field directions perfectly distinguishable.
In fact, Beckman \cite{beckman} has shown that for any three field vectors
forming a triangle that is isosceles or nearly isosceles, a suitable driving
field can always by found such that the field directions can be distinguished
perfectly.

\section{Distinguishing two alternatives}

Let's consider the special case in which our apparatus is known to be governed
by one of two possible Hamiltonians $H_1$ or $H_2$.  If the system is two
dimensional, we are trying to distinguish two possible values $\vec a,\vec b$
of the magnetic field with a spin-${1\over 2}$ probe.  Suppose for simplicity
that the two fields have the same magnitude (normalized to unity), but
differing directions.

Assuming that we are unable to modify the Hamiltonian by adding a driving term,
the optimal strategy is to choose an initial polarization vector that bisects
the two field directions $\hat a, \hat b$. Depending on the actual value of the
field, the polarization will precess on one of two possible cones. If the angle
between $\hat a$ and $\hat b$ is $\theta\ge 90^\circ$, then the two possible
polarizations will eventually be back-to-back; an orthogonal measurement
performed at that time will distinguish $\hat a$ and $\hat b$ perfectly. But if
$\theta < 90^\circ$, the two polarizations are never back-to-back; the best
strategy is to wait until the angle between the polarizations is maximal, and
to then perform the orthogonal measurement that best distinguishes them.  We
cannot perfectly distinguish the two field directions by this method.

On the other hand, if we are able to apply a known driving magnetic field in
addition to the unknown field that is to be determined, then two fields $\vec
a$ and $\vec b$ can always be perfectly distinguished.  If we apply the field
$-\vec b$, then the problem is one of distinguishing the trivial Hamiltonian
from
\begin{equation}
H_{\rm diff}=(\vec a - \vec b)\cdot \vec\sigma~.
\end{equation}
We can choose an initial polarization orthogonal to $\vec a - \vec b$, and wait
just long enough for $H_{\rm diff}$ to rotate the polarization by $\pi$.  Then
an orthogonal measurement perfectly distinguishes $H_{\rm diff}$ from the
trivial Hamiltonian.

Evidently, the same strategy can be applied to distinguish two Hamiltonians
$H_1$ and $H_2$  in a Hilbert space of arbitrary dimension. We drive the system
with $-H_2$; then to distinguish the trivial Hamiltonian from $H_1-H_2$, we
chose the initial state
\begin{equation}
{1\over\sqrt{2}}\left(|E_{\rm min}\rangle + |E_{\rm max}\rangle\right)~,
\end{equation}
where $E_{\rm min},E_{\rm max}$ are the minimal and maximal eigenvalues of
$H_1-H_2$.  After a time $t$ with
\begin{equation}
t(E_{\rm max}-E_{\rm min})=\pi~,
\end{equation}
this state evolves to the orthogonal state ${1\over\sqrt{2}}\left(|E_{\rm
min}\rangle - |E_{\rm max}\rangle\right)$, so that the trivial and nontrivial
Hamiltonians can be perfectly distinguished.

In the case of the two-dimensional version of the ``Grover problem'' with $H_1=
|0\rangle\langle 0|$ and $H_2=|1\rangle\langle 1 |$, this choice for the
driving Hamiltonian actually outperforms the Grover driving term of
Eq.~(\ref{grover_drive}) --- the two Hamiltonians can be distinguished in a
time that is shorter by a factor of $\sqrt{2}$.  So while the Grover strategy
is optimal for asymptotically large $N$, it is not actually optimal for $N=2$.

\section{Distinguishing two alternatives in a fixed time}

Let us now suppose that we are to distinguish between two time-independent
Hamiltonians $H_1$ and $H_2$, and that a {\sl fixed duration} $t$ has been
allotted to perform the experiment.  Is the driving strategy described above
(in which $-H_2$ is added to the Hamiltonian) always the best possible?

If we have the freedom to add a driving term of our choice, then we may assume
without loss of generality that we are to distinguish the nontrivial
Hamiltonian $H$ from the trivial Hamiltonian $0$. As already noted, if the
largest difference  $\Delta E=E_{\rm max}-E_{\rm min}$ of eigenvalues of $H$
satisfies $t\Delta E \ge \pi$, then $H$ can be perfectly distinguished from
$0$; let us therefore suppose that $t \Delta E  < \pi$.

If we add a {\sl time-independent} driving term $K$ to the Hamiltonian, and
choose an initial state $|\psi_0\rangle$, then after a time t, we will need to
distinguish the two states
\begin{equation}
\label{two_states}
e^{-i t K}|\psi_0\rangle~, \quad  e^{-it(H + K)}|\psi_0\rangle~.
\end{equation}
Two pure states will be more distinguishable when their inner product is
smaller. Therefore, to best distinguish $H+K$ from $K$, we should choose
$|\psi_0\rangle$ to minimize the inner product
\begin{equation}
\left|\langle\psi_0|e^{it K} e^{-it(H + K)}|\psi_0\rangle\right|~.
\end{equation}
If we expand $|\psi_0\rangle$ in terms of the eigenstates $\{|a\rangle\}$ of
$e^{it K} e^{-it(H + K)}$ with eigenvalues $\{e^{-itE_a}\}$,
\begin{equation}
|\psi_0\rangle=\sum_a \alpha_a|a\rangle~,
\end{equation}
this inner product becomes
\begin{equation}
\label{you_tee}
\left|\langle\psi_0|e^{it K} e^{-it(H + K)}|\psi_0\rangle\right|= \left|\sum_a
|\alpha_a|^2 e^{-itE_a}\right|~.
\end{equation}
The right-hand side of Eq.~(\ref{you_tee}) is the modulus of a convex sum of
points on the unit circle. Assuming the modulus is bounded from zero, it
attains its minimum when $|\psi_0\rangle$ is the equally weighted superposition
of the extremal eigenstates of $e^{it K} e^{-it(H + K)}$ -- those whose
eigenvalues are maximally separated on the unit circle.
For $K=0$, the minimum is
$\cos\left( t\Delta E/2\right)$, where $\Delta E$ is the difference of the
maximal and minimal eigenvalues of $H$.

We prove in Appendix A that turning on a nonzero driving term $K$ can never
cause the extremal eigenvalues to separate further, and therefore can never
improve the distinguishability of the two states in
Eq.~{\ref{two_states}.\footnote{That this might be the case was suggested to us
by Chris Fuchs \cite{fuchs_private}.} Therefore, $K=0$ is the optimal driving
term for distinguishing two Hamiltonians. In other words, if we wish to
distinguish between two Hamiltonians $H_1$ and $H_2$, it is always best to turn
on a driving term that precisely cancels one of the two.

The above discussion encompasses the strategy of introducing an ancilla
entangled with the probe (which proved effective for the problem of
distinguishing three or more alternatives). If we wish to distinguish two
Hamiltonians $H_1\otimes I$ and $H_2\otimes I$ that both act trivially on the
ancilla, the optimal driving term exactly cancels one of them ({\it e.g.}, $K=
- H_2\otimes I$), and so it too acts trivially on the ancilla. We derive no
benefit from the ancilla when there are only two alternatives.

Similarly, if we are trying to distinguish only two time-independent signals in
an allotted time, it seems likely there is no advantage to performing a
sequence of weak measurements, and adapting the driving field in response to
the incoming stream of measurement data.

\section{More alternatives: adaptive driving}

Now suppose that there are $N$ possible Hamiltonians ${H_1, H_2,
\ldots, H_N}$. If there is no time limitation, we can distinguish them
perfectly by implementing  an adaptive procedure; we make a series of
measurements, modifying our driving term and initial state in response to the
stream of measurement outcomes.

The correct Hamiltonian can be identified by pairwise elimination. First,
assume that either $H_1$ or $H_2$ is the actual Hamiltonian, and apply a
driving term to perfectly
distinguish them, say  $H_{D}=-H_1$.  After preparing the appropriate initial
state and waiting the appropriate time, we make an orthogonal measurement with
two outcomes --- the result indicates that either $H_1$ or $H_2$ is the
actual Hamiltonian.\footnote{Actually, in a Hilbert space of high dimension, we
can make a more complete measurement that will typically return the result that
neither $H_1$ nor $H_2$ is the actual Hamiltonian.}  If the result is $H_1$,
there are two possibilities:
either $H_1$ really is the Hamiltonian, or the assumption that one of $H_1$ or
$H_2$ is the Hamiltonian was wrong. Either way, $H_2$ has been eliminated.
Similarly, if
$H_2$ is found, $H_1$ is eliminated.  This procedure can then be repeated,
eliminating one Hamiltonian per measurement, thereby perfectly distinguishing
among the $N$ Hamiltonians in a total of $N-1$ measurements.

This algorithm is quite inefficient, however. The measurement record is $N-1$
bits long, while the information gain is only $\log N$ bits.

\section{Adaptive phase measurement and the semiclassical quantum Fourier
transform}

Far more efficient adaptive procedures can be formulated in some cases.
Consider, for example, a single qubit in a magnetic field of known direction
but unknown magnitude, so that
\begin{equation}
H_\omega= {\omega\over 2}\sigma_3~,
\end{equation}
and let us imagine that the value of the frequency $\omega$ is chosen
equiprobably from among $N=2^n$ equally spaced possible values. Without loss of
generality, we may normalize the field so that the possible values range from 0
to $1-2^{-n}$; then $\omega$ has a binary expansion
\begin{equation}
\omega = .\omega_1 \omega_2 \ldots \omega_n
\end{equation}
that terminates after at most $n$ bits.

The initial state $|\psi_0\rangle={1\over \sqrt{2}}(|0\rangle + |1\rangle)$
evolves in time $t$ to
\begin{equation}
|\psi(t)\rangle_\omega= e^{-itH_\omega}|\psi_0\rangle ={1\over
\sqrt{2}}(|0\rangle + e^{- i \omega t} |1\rangle)
\end{equation}
(up to an overall phase).  If we wait for a time $t_n=\pi 2^n$, the
final state is
\begin{equation}
|\psi(t_n)\rangle_\omega={1\over \sqrt{2}}(|0\rangle + e^{-i \pi \omega_n }
|1\rangle) ~.
\end{equation}
Now measurement in the $\{{1\over\sqrt{2}}(|0\rangle \pm |1\rangle\})$ basis
indicates (with certainty)
whether the bit $\omega_n$ is 0 or 1.  This outcome divides the set of possible
Hamiltonians in half, providing one bit of classical information.

The set of remaining possible Hamiltonians is still evenly spaced, but it may
have a constant offset, depending on the value of $\omega_n$. However, the
value of
$\omega_n$ is now known, so the offset can be eliminated. Specifically, if we
again prepare $|\psi_0\rangle$ and now evolve for a time $t_{n-1}=\pi
2^{n-1}$, we obtain the final state
\begin{equation}
|\psi(t_{n-1})\rangle_\omega={1\over \sqrt{2}}(|0\rangle + e^{-i \pi \>
(\omega_{n-1} . \omega_n)} |1\rangle)~.
\end{equation}
Since $\omega_n$ is known, we can perform a phase transformation (perhaps by
applying an additional driving magnetic field) to eliminate the phase $e^{-i
\pi \> (. \omega_n)}$;
Measuring again in the  $\{{1\over\sqrt{2}}(|0\rangle \pm |1\rangle\})$ basis
determines the value of $\omega_{n-1}$.

By continuing this procedure until all bits of $\omega$ are known, we perfectly
distinguish the $2^n$ possible Hamiltonians in just $n$ measurements.  The
procedure is optimal in the sense that we gain one full bit of information
about the Hamiltonian in each measurement.

Up until now we have imagined that the frequency $\omega$ takes one of $2^n$
equally spaced discrete values, but no such restriction is really necessary.
Indeed, what we have described is precisely the implementation of the $n$-qubit
semiclassical quantum Fourier transform as formulated by Griffiths and Niu
\cite{griffiths} (whose relevance to phase estimation was emphasized by Cleve
{\it et al.} \cite{cleve}).  Thus the same procedure can be applied to obtain
an estimate of the frequency to $n$-bit precision, even if the frequency is
permitted to take an arbitrary real value in the interval $[0,1)$.

Suppose that we attach to $n$ spins the labels $\{0,1,\ldots,n-2,n-1\}$, and
expose the $k$th spin to the field for time $\pi 2^{k+1}$; we thus prepare the
$n$-qubit state
\begin{equation}
\prod_{k=0}^{n-1} {1\over\sqrt{2}}\left(|0\rangle + e^{-i\pi\omega\cdot
2^{k+1}}|1\rangle\right)={1\over 2^{n/2}}\sum_{y=0}^{2^n-1} e^{-2\pi i \omega
\cdot y}|y\rangle~.
\end{equation}
The adaptive algorithm is equivalent to the quantum Fourier transform followed
by measurement;
hence the $n$-bit measurement outcome $\tilde \omega$ occurs with probability
\begin{equation}
{\rm Prob}_\omega(\tilde \omega) = \left| {1 \over 2^n} \sum_{y=0}^{2^n - 1}
\exp[-2 \pi i y (\omega-\tilde \omega)] \right|^2.
\end{equation}
If $\omega$ really does terminate in $n$ bits, then the outcome $\tilde\omega$
is
guaranteed to be its correct binary expansion.  But even if the binary
expansion of $\omega$ does not terminate, the probability that our estimate
$\tilde \omega$ is correct to $n$ bits of precision is still of order
one.\footnote{We might also use the QFT to {\sl compute} eigenvalues of a known
many-body Hamiltonian, rather than {\sl measure} eigenvalues of an unknown one
\cite{lloyd}.}

Of course, to measure the frequency to a precision $\Delta \omega$ of order
$2^{-n}$, we need to expose our probe spins to the unknown Hamiltonian for a
total time $T$ of order $2\pi\cdot 2^{n}$. The accuracy is limited by an
energy-time uncertainty relation of the form $T\Delta\omega\sim 1$.

The semiclassical quantum Fourier transform provides an elegant solution to the
problem of performing an ideal ``phase measurement'' in the Hilbert space of
$n$ qubits.  More broadly, any $N$-dimensional Hilbert space with a preferred
basis $\{|k\rangle, ~k=0,1,\dots, N-1\}$ has a complementary basis of {\sl
phase states}
\begin{equation}
|\varphi\rangle={1\over\sqrt{N}} \sum_{k=o}^{N-1}e^{ik\varphi}|k\rangle~,
\end{equation}
with
\begin{equation}
\varphi = 2\pi j/N~,\quad j=0,1,\dots,N-1~.
\end{equation}
For example, the Hilbert space could be the truncated space of a harmonic
oscillator like a mode of the electromagnetic field, with the occupation number
restricted to be less than $N$; then the states $|\varphi\rangle$ are the
``phase squeezed'' states of the oscillator that have minimal phase
uncertainty. Since a POVM in an $N$-dimensional Hilbert space can acquire no
more than $\log N$ bits of information about the preparation of the quantum
state, the phase of an oscillator with occupation number less than $N$ can be
measured to at best $\log N$ bits of accuracy. While it is easy to do an
orthogonal measurement in the occupation number basis with an efficient
photodetector, an orthogonal measurement in the $|\varphi\rangle$ basis is
quite difficult to realize in the laboratory \cite{wiseman}.

But if the standard basis is the computational basis in the $2^n$-dimensional
Hilbert space of $n$ qubits, then an ideal phase measurement is simple to
realize.  Since the phase eigenstates are actually not entangled states, we can
carry out the measurement -- {\sl adaptively} -- one qubit at a time.

Note that if we had an arbitrarily powerful quantum computer with an
arbitrarily large amount of quantum memory, then adaptive measurement
strategies might seem superfluous.  We could achieve the same effect by
introducing a large ancilla and a driving Hamiltonian that acts on probe and
ancilla, with all measurements postponed to the very end.  But the
semiclassical quantum Fourier transform illustrates that adaptive techniques
can reduce the complexity of the quantum information processing required to
perform the measurement.  In many cases, an adaptive strategy may be realizable
in practice, while the equivalent unitary strategy is completely infeasible.

\section{Distinguishability and decoherence}

In all of our examples so far, we have ignored noise and decoherence.  In
practice, decoherence may compromise our ability to decipher the classical
signal with high confidence.  Finding ways to improve measurement accuracy by
effectively coping with decoherence is an important challenge faced by quantum
information theory.

If there is decoherence, our aim is to gain information about the value of a
parameter in a master equation rather than a Hamiltonian.  To be concrete,
consider a single qubit governed by an unknown Hamiltonian $H$, and also
subject to decoherence described by the ``depolarizing channel;''  the density
matrix $\rho$ of the qubit obeys the master equation
\begin{equation}
\dot \rho= -i[H,\rho] - \Gamma\left(\rho-{1\over 2} I\right)~,
\end{equation}
where $\Gamma$ is the (known) damping rate.
If we express $\rho$ in terms of the polarization vector $\vec P$,
\begin{equation}
\rho={1\over 2}(I+\vec P\cdot\vec\sigma)~,
\end{equation}
and the Hamiltonian as
\begin{equation}
H={\omega\over 2}~\hat a\cdot \sigma~,
\end{equation}
then the master equation becomes
\begin{equation}
\dot{\vec P}=\omega(\hat a \times \vec P) - \Gamma \vec P~.\
\end{equation}
The polarization precesses uniformly with circular frequency $\omega$ about the
$\hat a$-axis as it contracts with lifetime $\Gamma^{-1}$.

Suppose that we are to distinguish among two possible Hamiltonians, which are
assumed to be equiprobable. If we are able to add a driving term, we may assume
that the two are the trivial Hamiltonian and
\begin{equation}
H={\omega\over 2}~\sigma_3~.
\end{equation}
We choose the initial polarization vector $P_0=(1,0,0)$.  Then if the
Hamiltonian is trivial, the polarization contracts as
\begin{equation}
\vec P(t)_{\rm triv}= e^{-\Gamma t}(1,0,0)~,
\end{equation}
while under the nontrivial Hamiltonian it contracts and rotates as
\begin{equation}
\vec P(t)_{\rm nontriv}=e^{-\Gamma t}(\cos \omega t,\sin \omega t, 0)~.
\end{equation}

When is the best time to measure the polarization? We should wait until $\vec
P_{\rm triv}$ and $\vec P_{\rm nontriv}$ point in distinguishable directions,
but if we wait too long, the states will depolarize.  The optimal measurement
to distinguish the two is an orthogonal measurement of the polarization along
the axis normal to the bisector of the vectors $\vec P(t)_{\rm triv}$ and $\vec
P(t)_{\rm nontriv}$.  At time $t$ the probability that this measurement
identifies the Hamiltonian incorrectly is
\begin{equation}
P_{\rm error}= {1\over 2} - {1\over 2} e^{-\Gamma t}\left|\sin\left({\omega
t\over 2}\right)\right|~.
\end{equation}
This error probability is minimized, and the information gain from the
measurement is maximized, at a time $t$ such that
\begin{equation}
\tan\left({\omega t\over 2}\right)= {\omega\over 2\Gamma}~.
\end{equation}
If $\Gamma/\omega<<1$, this time is close to $\pi/\omega$, the time we would
measure to perfectly distinguish the Hamiltonians in the absence of
decoherence.  But if $\Gamma/\omega >>1$, then we should measure after a time
$t\sim \Gamma^{-1}$ comparable to the lifetime.

More generally, consider an ensemble of two density operators $\rho_1$ and
$\rho_2$ with {\it a priori}} probabilities $p_1$ and $p_2$ (where $p_1 +
p_2=1$), and imagine that an unknown state has been drawn from this ensemble. A
procedure for deciding whether the unknown state is $\rho_1$ or $\rho_2$ can be
modeled as a POVM with two outcomes. The two-outcome POVM that minimizes the
probability of making an incorrect decision is a measurement of the orthogonal
projection onto the space spanned by the eigenstates of $p_1\rho_1-p_2 \rho_2$
with positive eigenvalues \cite{helstrom,fuchs_thesis}. The minimal error
probability achieved by this measurement is
\begin{equation}
P_{\rm error}={1\over 2}- {1\over 2}{\rm tr}\left|p_1\rho_1-p_2\rho_2\right|~.
\end{equation}

Correspondingly, if we are to identify an unknown superoperator as one of
$\$_1$ and $\$_2$ (with {\it a priori} probabilities $p_1$ and $p_2$), then the
way to distinguish $\$_1,\$_2$ with minimal probability of error is to choose
our initial state $\rho_0=|\psi_0\rangle\langle \psi_0|$ to
minimize\footnote{We thank Chris Fuchs for a helpful discussion of this point.}
\begin{equation}
\label{super_error}
P_{\rm error}={1\over 2}- {1\over 2}{\rm
tr}\left|\left(p_1\$_1-p_2\$_2\right)\rho_0\right|~.
\end{equation}
In the case of interest to us, the superoperators $\$_1$ and $\$_2$ are
obtained my integrating, for time $t$, master equations with Hamiltonians $H_1$
and $H_2$ respectively. We minimize the error probability in
Eq.~(\ref{super_error}) with respect to $t$ to complete the optimization.

\section{Entanglement and frequency measurement}

Consider again the case in which the Hamiltonian is known to be of the form
\begin{equation}
H_{\omega}={\omega\over 2}~\sigma_3~,
\end{equation}
but where the frequency $\omega$ is unknown.  For the moment, let us neglect
decoherence, but suppose that we have been provided with a large number $n$ of
qubits that we may use to perform an experiment to determine $\omega$ in a {\sl
fixed total time} $t$.  What is the most effective way to employ our qubits?

Consider two strategies.  In the first, we prepare $n$ identical qubits
polarized along the $x$-axis.  They precess in the field described by
$H_\omega$ for time $t$, and then the spin along the $x$-axis is measured. Each
spin will be found to be pointing ``up'' with probability
\begin{equation}
P = {1\over 2}(1+ \cos\omega t)
\end{equation}
Because the measurement is repeated many times, we will be able to estimate the
probability $P$ to an accuracy
\begin{equation}
\label{n_qubits}
\Delta P=\sqrt{P(1-P)/n}={|\sin\omega t|\over 2\sqrt{n}}~.
\end{equation}
and so determine the value of $\omega$ to accuracy
\begin{equation}
\label{shot_noise}
\Delta \omega = {\Delta P\over t |dP/d(\omega t)|}={1\over t \sqrt n}~.
\end{equation}
The accuracy improves like $1/\sqrt{n}$ as we increase the number of available
qubits with the time $t$ fixed.

The second strategy is to prepare an entangled ``cat'' state of $n$ ions
\begin{equation}
|\psi_0\rangle = {1\over \sqrt{2}}(|000\dots0\rangle + |111\dots 1\rangle)~.
\end{equation}
The advantage of the entangled state is that it precesses $n$ times faster than
a single qubit; in time $t$ it evolves to
\begin{equation}
|\psi(t)\rangle= {1\over \sqrt{2}}(|000\dots0\rangle + e^{i n\omega t}|111\dots
1\rangle)
\end{equation}
(up to an overall phase).
If we now perform an orthogonal measurement that projects onto the basis
${1\over \sqrt{2}}(|000\dots0\rangle \pm |111\dots 1\rangle)$ ({\it e.g.} a
measurement of the entangled observable
$\sigma_1\otimes\sigma_1\otimes\cdots\otimes\sigma_1$) then we will obtain the
``+'' outcome with probability
\begin{equation}
P={1\over 2}(1+\cos n\omega t)~.
\end{equation}
By this method, $n\omega t$ can be measured to order one accuracy, so that
\begin{equation}\
\label{linear_noise}
\Delta\omega\simeq { 1\over tn}~,
\end{equation}
a more favorable scaling with $n$ than in Eq.~(\ref{shot_noise}).

This idea of exploiting the rapid precession of entangled states to achieve a
precision beyond the shot-noise limit has been proposed in both frequency
measurement \cite{wineland} and optical interferometry \cite{yurke}. (One
realization of this idea is the proposal by Caves \cite{caves} to allow a
squeezed vacuum state to enter the dark port of an interferometer; the
squeezing induces the $n$ photons entering the other port to make correlated
``decisions'' about which arm of the interferometer to follow.)

\section{Entanglement versus decoherence}
In both Eq.~(\ref{shot_noise}) and Eq.~(\ref{linear_noise}), the accuracy of
the frequency measurement improves with the elapsed time $t$ as $1/t$.  But so
far we have neglected decoherence.  If the single-qubit state decays at a rate
$\Gamma$, then we have seen that the optimal time at which to perform a
measurement will be of order $\Gamma^{-1}$.  The entangled strategy will still
be better if we are constrained to perform the measurement in a time
$t<<\Gamma^{-1}$, but further analysis is needed to determine which method is
better if we are free to choose the time $t$ to optimize the accuracy.

In fact, as Huelga {\it et al.} \cite{huelga} have emphasized, an entangled
state is fragile, and its faster precession can be offset by its faster decay
rate. Suppose that two qubits are available, both independently subjected to
the depolarizing channel with decay rate $\Gamma$.
If we prepare the unentangled state, each qubit has the initial pure-state
density matrix
\begin{equation}
\rho_0={1\over 2}(I+\sigma_1)~
\end{equation}
polarized along the $x$-axis, and evolves in time $t$ to
\begin{equation}
\rho(t)={1\over 2}[I+e^{-\Gamma t}(\sigma_1~\cos\omega t+ \sigma_2~\sin\omega
t)]~.
\end{equation}
If we now measure $\sigma_1$, we obtain the $+$ result with probability
\begin{equation}
\label{nonentangle_prob}
\label{single_decohere}
P={\rm tr}\left({1\over 2}(I+\sigma_1)\rho(t)\right)={1\over 2}(1+e^{-\Gamma
t}\cos\omega t)~.
\end{equation}

Now suppose that the initial state is the Bell state $|\phi^+\rangle$ of two
qubits, with density matrix
\begin{equation}
\rho_0={1\over 4}\left(I\otimes I+ \sigma_3\otimes\sigma_3 +
\sigma_1\otimes\sigma_1-\sigma_2\otimes\sigma_2\right)~.
\end{equation}
If both spins precess and depolarize independently, this state evolves to
\begin{eqnarray}
\rho(t)& = & {1\over 4} [I\otimes I+ e^{-2\Gamma t}\big(\sigma_3\otimes\sigma_3
\nonumber\\
& + &\cos 2\omega t(\sigma_1\otimes\sigma_1-\sigma_2\otimes\sigma_2)
\nonumber\\
& + & \sin 2\omega t(\sigma_1\otimes\sigma_2+\sigma_2\otimes\sigma_1)\big)]~;
\end{eqnarray}
if we measure the observable $\sigma_1\otimes\sigma_1$, we find the + outcome
with probability
\begin{eqnarray}
\label{entangle_prob}
P & = &{\rm tr}\left({1\over 2}(I\otimes I
+\sigma_1\otimes\sigma_1)\rho(t)\right)\nonumber \\
& = & {1\over 2}(1+e^{-2\Gamma t}\cos2 \omega t)~.
\end{eqnarray}

Note that Eq.~(\ref{entangle_prob}) has exactly the same functional form as
Eq.~(\ref{nonentangle_prob}), but with $t$ replaced by $2t$.  Therefore, the
entangled measurement performed in time $t/2$ collects exactly as much
information about the frequency $\omega$ as the measurement of a single ion
performed in time $t$.  If we have two qubits and total time $t$ available, we
can either perform the entangled measurement twice (taking time $t/2$ each
time), or perform measurements on each qubit independently (taking time $t$).
Either way, we obtain two outcomes and collect exactly the same amount of
information on the average.

More generally, suppose that we have $n$ qubits and a total time $T>> 1/
\Gamma$ available.  We can use these qubits to perform altogether $nT/t$
independent single-qubit measurements, where each measurement requires time
$t$. Plugging Eq.~(\ref{single_decohere}) into Eq.~(\ref{n_qubits}) and
Eq.~(\ref{shot_noise}) (with $n$ replaced by $nT/t$), and choosing $\cos\omega
t\sim 0$ to optimize the precision, we find that the frequency can be
determined to accuracy
\begin{equation}
\Delta\omega=\left({1\over t}\right)\cdot {e^{\Gamma t}\over
\sqrt{nT/t}}={1\over \sqrt{nT}}\cdot {e^{\Gamma t}\over \sqrt{t}}~.
\end{equation}
This precision is optimized if we choose $\Gamma t=1/2$, where we obtain
\cite{huelga}
\begin{equation}
\Delta\omega = \sqrt{2e\Gamma\over nT}~.
\end{equation}
On the other hand, we could repeat the experiment $T/t$ times using the
$n$-qubit entangled state.  Then we would obtain a precision
\begin{equation}
\Delta\omega = \left({1\over nt}\right)\cdot {e^{n\Gamma t}\over
\sqrt{T/t}}={1\over \sqrt{nT}}\cdot {e^{n\Gamma t}\over \sqrt{nt}}~,
\end{equation}
the same function as for uncorrelated qubits, but with $t$ replaced by $nt$.
Thus the optimal precision is the same in both cases, but is attained in the
uncorrelated case by performing experiments that take $n$ times longer than in
the correlated case.

That the entangled states offer no advantage in the determination of $\omega$
was one of the main conclusions of Huelga {\it et al.} \cite{huelga}.  A
similar conclusion applies to estimating the difference in path length between
two arms of an interferometer using a specified optical power, if we take into
account losses and optimize with respect to the number of times the light
bounces inside the interferometer before it escapes and is detected.

We would like to make the (rather obvious) point that this conclusion can
change if we adopt a different model of decoherence, and in particular if the
qubits do not decohere independently. As a simple example of correlated
decoherence, consider the case of two qubits with $4 \times 4$ density matrix
$\rho$ evolving according to the master equation
\begin{equation}
\dot \rho = -i[H,\rho] - \Gamma \left(\rho-I/4\right)~.
\end{equation}
This master equation exhibits the analog, in the four-dimensional Hilbert
space, of the uniform contraction of the Bloch sphere described by the
depolarizing channel in the case of a qubit.  Because the decoherence picks out
no preferred direction in the Hilbert space (or any preferred tensor-product
decomposition), we call this model ``symmetric decoherence.''

Under this master equation, with both qubits subjected to $H_\omega$ and to
symmetric decoherence, the Bell state $\rho_0=|\phi^+\rangle\langle\phi^+|$
evolves in time $t$ to the state
\begin{eqnarray}
\rho(t)& = & {1\over 4} [I\otimes I+ e^{-\Gamma t}\big(\sigma_3\otimes\sigma_3
\nonumber\\
& + &\cos 2\omega t(\sigma_1\otimes\sigma_1-\sigma_2\otimes\sigma_2)
\nonumber\\
& + & \sin 2\omega t(\sigma_1\otimes\sigma_2+\sigma_2\otimes\sigma_1)\big)]~,
\end{eqnarray}
so that a measurement of $\sigma_1\otimes\sigma_1$ yields the + outcome with
probability
\begin{equation}
\label{ent_symmetric}
P ={1\over 2}(1+e^{-\Gamma t}\cos2 \omega t)~.
\end{equation}
On the other thing, the initial product state
\begin{equation}
\rho_0={1\over 4} (I+\sigma_1)\otimes(I+\sigma_1)
\end{equation}
becomes entangled as a result of symmetric decoherence. Were the Hamiltonian
trivial, it would evolve to
\begin{equation}
\rho(t)={1\over 4} I\otimes I + {1\over 4}e^{-\Gamma t}(\sigma_1\otimes I+
I\otimes\sigma_1 +\sigma_1\otimes\sigma_1)~.
\end{equation}
Including the precession
\begin{equation}
\sigma_1\to \sigma_1 \cos\omega t + \sigma_2\sin\omega t~,
\end{equation}
we obtain
\begin{equation}
\rho(t)={1\over 4} I\otimes I + {1\over 4}e^{-\Gamma t}( \sigma_1\otimes
I~\cos\omega t + \cdots~)~,
\end{equation}
so that measurement of the single-qubit observable $\sigma_1\otimes I$ yields
the + outcome with probability
\begin{equation}
\label{single_symmetric}
P={\rm tr}\left({1\over 2}(I\otimes I +\sigma_1\otimes I)\rho(t)\right)={1\over
2}(1+e^{-\Gamma t}\cos\omega t)~.
\end{equation}
Comparing Eq.~(\ref{single_symmetric}) and Eq.~(\ref{ent_symmetric}), the
important thing to notice is that with symmetric decoherence, entangled states
decay no faster than product states; therefore, we can enjoy the benefit of
entanglement (faster precession) without paying the price (faster decay).

To establish more firmly that entangled strategies outperform nonentangled
strategies in the symmetric decoherence model, we should consider more closely
what are the optimal final measurements for these two types of initial states.
To give the problem a precise information-theoretic formulation, we return to
the problem of distinguishing two cases, the trivial Hamiltonian and
$H_\omega$, which are assumed to be equiprobable.  For either the product
initial state or the entangled initial state, we evolve for time $t$, and then
perform the best measurement that distinguishes between evolution governed by
$H_\omega$ and trivial evolution. In both cases, the measurement is permitted
to be an entangled measurement; that is, we optimize with respect to all POVM's
in the four-dimensional Hilbert space.

In either case (initial product state or initial entangled state), we can find
the two-outcome POVM that identifies the Hamiltonian with minimal probability
of error. When there is no decoherence,  this POVM (when restricted to the
two-dimensional subspace containing the two pure states to be distinguished) is
the familiar orthogonal measurement that best distinguishes two pure states of
a qubit.  In fact, for symmetric decoherence, this same measurement minimizes
the error probability for any value of the damping rate $\Gamma$.  It is thus
the two-outcome measurement with the maximal information gain  (the measurement
outcome has maximal mutual information with the choice of the Hamiltonian).
Although we don't have a proof, we can make a reasonable guess that, for
symmetric decoherence, this two-outcome measurement has the maximal information
gain of any measurement, including POVM's with more outcomes.

If either initial state evolves for time $t$, and then this optimal POVM is
performed, the error probability can be expressed as
\begin{equation}
P_{\rm error} = {1\over 2} - {1\over 2}e^{-\Gamma
t}\left|\sin\theta(t)\right|~;
\end{equation}
here $\theta(t)$ is the angle between the states --- that is, $\cos \theta(t)$
is the inner product of the evolving and static states, in the limit of no
damping ($\Gamma=0$). For the entangled initial state, we have
\begin{equation}
\theta_{\rm entangled}=\omega t~,
\end{equation}
and for the product initial state, we have
\begin{equation}
\cos\theta_{\rm product}=\cos^2 \left({\omega t\over 2}\right)~.
\end{equation}
Since
\begin{equation}
|\cos \theta_{\rm entangled}| = |\cos\omega t| \le {1\over 2}(1+\cos \omega t)
=|\cos\theta_{\rm product}|
\end{equation}
for $\cos\theta_{\rm entangled}\ge 0$, the error probability achieved by the
entangled initial state is smaller than that achieved by the product state for
$0 < \omega t< \pi/2$, which is sufficient to ensure that the error probability
optimized with respect to $t$ is always smaller in the entangled case for any
nonzero value of $\Gamma$.
Similarly, if we optimize the information gain with respect to $t$, the
entangled strategy has the higher information gain for all $\Gamma>0$.  The
improvement in information gain (in bits) achieved using an entangled initial
state rather than a product initial state is plotted in Fig.~1 as a function of
$\Gamma/\omega$. The maximum improvement of about .136 bits occurs for
$\Gamma/\omega\sim .379$.

\begin{figure}[t]
\begin{center}
\leavevmode
\epsfxsize=3in
\epsfbox{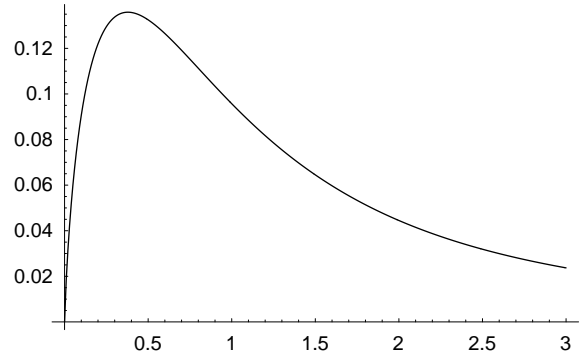}
\end{center}
\caption{Improvement in information gain (in bits) achieved by using an
entangled initial state, as a function of the ratio of decoherence rate
$\Gamma$ to precession frequency $\omega$.}
\end{figure}

We have already seen in \S II that, even in the absence of decoherence, an
entangled strategy may outperform an unentangled strategy if we are trying to
distinguish more than two alternatives.  This advantage will persist when
sufficiently weak decoherence is included, whether correlated or uncorrelated.
In that event, since only one member of an entangled pair is exposed to the
unknown Hamiltonian, we may be able to shelter the other member of the pair
from the ravages of the environment, slowing the decay of the state and
strengthening the signal.

\section{Conclusions}
We feel that quantum information theory, having already secured a central
position at the foundations of computer science, will eventually erect bridges
connecting with many subfields of physics. The results reported here (and other
related examples) give strong hints that ideas emerging from the theory of
quantum information and computation are destined to profoundly influence the
experimental physics techniques of the future.

We have only scratched the surface of this important subject.  Among the many
issues that deserve further elaboration are the connections between
superoperator distinguishability and superoperator norms, the efficacy of the
quantum Fourier transform in the presence of decoherence, the measurement of
continuous quantum variables, the applications of quantum error correction, and
the detection of time-dependent signals.

\acknowledgments

We thank Constantin Brif, Jon Dowling, Steven van Enk, Jeff Kimble, Alesha
Kitaev, and Kip Thorne for instructive discussions about quantum measurement.
We are especially grateful to Hideo Mabuchi for introducing us to this
fascinating subject, to Chris Fuchs for sharing his insights into state
distinguishability, and to Dave Beckman for discussions on improving the
superdense coding method by applying a driving field. Thanks to Barry Simon for
useful comments on the theorem in Appendix A, and for persuading us that it is
not completely trivial. We also thank  C. Woodward for helpful correspondence.
A.~M.~C. and J.~R. received support from Caltech's Summer Undergraduate
Research Fellowship (SURF) program, and A.~M.~C. received a fellowship endowed
by Arthur R. Adams. This work has been supported in part by the Department of
Energy under Grant
No. DE-FG03-92-ER40701, and by DARPA through the Quantum Information and
Computation (QUIC) project administered by the Army Research Office under Grant
No. DAAH04-96-1-0386.

\section*{Appendix A: Fixed-time-driving theorem}
In this appendix, we sketch the proof of the theorem stated in \S V.

For a unitary $N\times N$ matrix $U$, we define ${\rm maxarg}(U)$ to be the
largest argument of an eigenvalue of $U$, where the argument takes values in
the interval $(-\pi,\pi]$.  Similarly, ${\rm minarg}(U)$ is the minimum
argument of an eigenvalue of $U$. Our theorem is:\smallskip

{\bf Theorem 1.} {\sl If $H$ and $K$ are finite-dimensional Hermitian matrices,
and $\parallel H\parallel_{\rm sup} <\pi$, then
\begin{eqnarray}
\label{thm1}
{\rm maxarg}\left(e^{iK} e^{-i(H+K)}\right) & \le & {\rm maxarg}(e^{-iH}) ~,\\
\label{thm2}
{\rm minarg}\left(e^{iK} e^{-i(H+K)}\right) & \ge & {\rm minarg}(e^{-iH})~.
\end{eqnarray}
}

To prove the theorem, we begin with:\smallskip

{\bf Lemma 2}. {\sl For unitary $U$ with ${\rm maxarg}(U) \ne \pi$, and
Hermitian $A$,
\begin{eqnarray}
\label{maxlemma2}
& &{\rm maxarg} (Ue^{i\varepsilon A})\le    {\rm maxarg}(U) + {\rm
maxarg}(e^{i\varepsilon A})+ O(\varepsilon^2)~,\nonumber\\
& & \\
\label{minlemma2}
& &{\rm minarg} (Ue^{i\varepsilon A})\ge    {\rm minarg}(U) + {\rm
minarg}(e^{i\varepsilon A})- O(\varepsilon^2)~.\nonumber\\
& &
\end{eqnarray}
}

\noindent {\sl Proof}: Write $U=e^{iB}$, where $B$ is Hermitian and
\newline $\parallel B \parallel_{\rm sup} < \pi$; then maxarg$(e^{iB})={\rm
max}(B)$, where ${\rm max}(B)$ denotes the maximum eigenvalue of $B$. From the
Baker-Campbell-Hausdorff formula, we have
\begin{equation}
e^{iB}e^{i\varepsilon A}= \exp i\left(B+\varepsilon A + {i\over 2}\varepsilon
[C,B] + O(\varepsilon^2)\right) ~,
\end{equation}
where $C$ is linear in $A$. Then lowest-order eigenvalue perturbation theory
tells us that
\begin{eqnarray}
& &{\rm max}\left(B+\varepsilon A + {i\over 2}\varepsilon
[C,B]\right)\nonumber\\
&=& {\rm max}(B) +\langle \psi|\left(\varepsilon A + {i\over 2}\varepsilon
[C,B]\right)|\psi\rangle +O(\varepsilon^2)\nonumber\\
&=&{\rm max}(B)+\langle\psi|\left(\varepsilon A
\right)|\psi\rangle+O(\varepsilon^2)\nonumber\\
&\le&{\rm max}(B) +{\rm max}(\varepsilon A)+O(\varepsilon^2)
\end{eqnarray}
(where $|\psi\rangle$ is in the eigenspace of $B$ with maximal eigenvalue).
This proves Eq.~(\ref{maxlemma2}). Eq.~(\ref{minlemma2}) is proved similarly.
Note that the condition ${\rm maxarg(U)}\ne\pi$ is necessary so that the
singularity of the maxarg function can be avoided for $\varepsilon$
sufficiently small.

Lemma 2 is all we will need for the proof of Theorem 1. But it is useful to
note that Lemma 2 may be invoked to prove:\smallskip

{\bf Lemma 3}.\footnote{Strangely, we could find only one reference to this
proposition in the literature; it is a special case of Eq.~(8) in
\cite{woodward}.} {\sl For unitary $U_1$ and $U_2$, such that
\begin{eqnarray}
\label{maxargcond}
{\rm maxarg}(U_1) + {\rm maxarg}(U_2) &<& \pi~,\\
\label{minargcond}
{\rm minarg}(U_1) + {\rm minarg}(U_2) &>& -\pi ~,
\end{eqnarray}
we have
\begin{eqnarray}
\label{maxarg}
{\rm maxarg}(U_1 U_2) & \le & {\rm maxarg}(U_1) + {\rm maxarg}(U_2) ~,\\
\label{minarg}
{\rm minarg}(U_1 U_2) & \ge & {\rm minarg}(U_1) + {\rm minarg}(U_2)~.
\end{eqnarray}
}

\noindent{\sl Proof}: We write
\begin{equation}
U_1U_2 = U_1 e^{iA}= U_1 \left(e^{iA/n}\right)^n~,
\end{equation}
where the eigenvalues of A lie in the interval $(-\pi,\pi)$, and apply Lemma 2
repeatedly, obtaining
\begin{eqnarray}
{\rm maxarg}\left(U_1e^{iA}\right)&\le& {\rm maxarg}(U_1)\nonumber\\
&+& n\left[{\rm maxarg}(e^{iA/n})+ O(n^{-2})\right]~.
\end{eqnarray}
Taking the $n\to\infty$ limit proves Eq.~(\ref{maxarg}). Eq.~(\ref{minarg}) is
proved similarly.
Note that because of the conditions Eq.~(\ref{maxargcond}) and
Eq.~(\ref{minargcond}), Lemma 2 can be safely applied $n$ times in succession;
the accumulated maxarg and minarg of the product never approach $\pi$.

To complete the proof of Theorem 1, we invoke the Lie product formula
\begin{equation}
\lim_{n \to \infty} (e^{A/n} e^{B/n})^n = e^{A+B}~,
\end{equation}
to write
\begin{eqnarray}
\label{expexpand}
&&e^{iK}e^{-i(H+K)} = \lim_{n \to \infty} (e^{iK/n})^n (e^{-iH/n}
                        e^{-iK/n})^n \nonumber \\
                  & = & \lim_{n \to \infty} e^{iK/n} \cdots e^{iK/n}
                        e^{-iH/n} e^{-iK/n} \cdots e^{-iH/n} e^{-iK/n}~.
\end{eqnarray}
Since $e^{iK/n} e^{-iH/n} e^{-iK/n}$ and $e^{-iH/n}$ have the same eigenvalues,
Lemma 3 implies that
\begin{eqnarray}
&{\rm maxarg}&(e^{iK/n} e^{-iH/n} e^{-iK/n} e^{-iH/n}) \nonumber\\
     & \le & 2 \cdot {\rm maxarg}(e^{-iH/n}) ~.
\end{eqnarray}
Similarly, we have
\begin{eqnarray}
&{\rm maxarg}&\left(e^{iK/n}\left(e^{iK/n} e^{-iH/n} e^{-iK/n} e^{-iH/n}\right)
e^{-iK/n}e^{-iH/n}\right) \nonumber\\
     & \le & 3 \cdot {\rm maxarg}(e^{-iH/n}) ~,
\end{eqnarray}
and so on.  Hence, applying Lemma 3 altogether $n$ times to the right-hand side
of Eq.~(\ref{expexpand}), we find that
\begin{eqnarray}
{\rm maxarg}\left(e^{iK} \left(e^{-iH/n} e^{-iK/n}\right)^n\right)
     & \le & n \cdot {\rm maxarg}\left(e^{-iH/n}\right) \nonumber\\
&= &{\rm maxarg}(e^{-iH}) \\
\end{eqnarray}
Taking the $n\to\infty$ limit completes the proof of Eq.~(\ref{thm1}).
Eq.~(\ref{thm2}) is proved similarly.

The upper bound on $\parallel H\parallel_{\rm sup}$ is a key feature of the
formulation of Theorem 1.  This bound ensures that the conditions
Eq.~(\ref{maxargcond}) and Eq.~(\ref{minargcond}) are satisfied each time that
Lemma 3 is invoked in the proof. If $\parallel H\parallel_{\rm sup}$ is too
large, then counterexamples can be constructed.

In any event, for the discussion in \S V, we are interested in the case where
the maximal and minimal eigenvalues of $H$ differ by less than $\pi$, and by
shifting $H$ by a constant we can ensure that $\parallel H\parallel_{\rm sup}<
\pi/2$.  Therefore, the theorem enforces the conclusion that if we are to
distinguish a nontrivial Hamiltonian from the trivial Hamiltonian in an
experiment conducted in a fixed elapsed time, turning on a nonzero
time-independent ``driving term'' $K$ provides no advantage.

\end{document}